\documentclass[aps,prl,superscriptaddress,showkeys]{revtex4}
\usepackage{epsfig}
\begin{document}

\title{Imaging with Scattered Neutrons}

\author{H. Ballhausen}
\affiliation{Institute of Physics, University of Heidelberg, Philosophenweg 12, 69120 Heidelberg, Germany}
\email{ballhausen@physi.uni-heidelberg.de}

\author{H. Abele}
\affiliation{Institute of Physics, University of Heidelberg, Philosophenweg 12, 69120 Heidelberg, Germany}

\author{R. G\"ahler}
\affiliation{Institut Laue-Langevin, 6 rue Jules Horowitz BP 156, 38042 Grenoble Cedex 9, France}

\author{M. Trapp}
\affiliation{Institute of Physics, University of Heidelberg, Philosophenweg 12, 69120 Heidelberg, Germany}

\author{A. Van Overberghe}
\affiliation{Institut Laue-Langevin, 6 rue Jules Horowitz BP 156, 38042 Grenoble Cedex 9, France}

\begin{abstract}
\noindent

	We describe a novel experimental technique for neutron imaging with scattered neutrons.
	These scattered neutrons are of interest for condensed matter physics, because they
	permit to reveal the local distribution of incoherent and coherent scattering within a sample.
	In contrast to standard attenuation based imaging, scattered neutron imaging distinguishes
	between the scattering cross section and the total attenuation cross section including absorption.
	First successful low-noise millimeter-resolution images by scattered neutron radiography and tomography are presented.

\end{abstract}


\maketitle

\section{Introduction}


Conventional neutron imaging \cite{Kallmann1948,ITMNR5} always uses a coaxial setup of neutron source,
sample and detector (see Fig. 1).
The neutron beam is attenuated by absorption and scattering,
its shadow being cast on a position sensitive detector behind the sample.
The projections obtained in this way are then used to produce radiographs and
tomographs. Consequently, these images visualize the total attenuation
cross section, the sum of both absorption and scattering cross sections.

\bigskip
\noindent
\begin{minipage}[h]{7cm}
~~The ratio of absorption to scattering varies greatly between different
elements. It is therefore crucial to be able to separate the contributions of
scattering and absorption or to visualize even coherently scattering inclusions
in attenuating bulk material. Many experiments like 3D crystallographic imaging,
 strain scanning and on hydrogen storage will benefit from and require this basic capability.
Scattered neutron radiography has been investigated \cite{Kobayashi2002}, however,
the results were described as preliminary, demanding significant improvements of the technique. 
\end{minipage}
~~~~~~~~~~
\begin{minipage}[h]{10cm}
\begin{center}
\epsfig{file=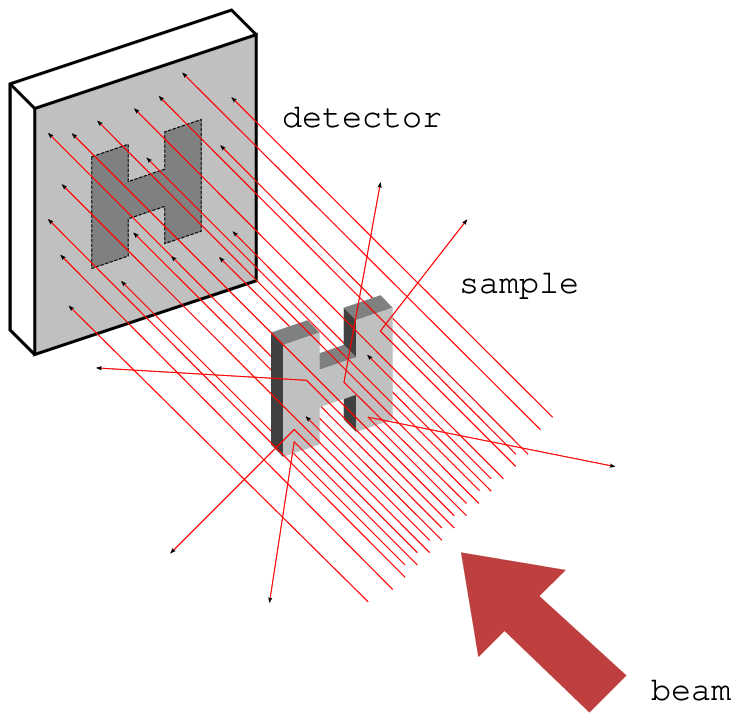,height=4.1cm}\\
Fig. 1: Conventional neutron imaging:\\
the detector, placed on the beam axis, views the\\
beam attenuated by both absorption and scattering.
\end{center}
\end{minipage}

~

\bigskip
\noindent
\begin{minipage}[h]{7cm}
~~Here, we describe a novel, off-angular setup (see Fig. 2).
The detector is set far off the beam axis and views the sample under a significant angle. 
Only scattered neutrons are detected in this way. Position sensitivity is created by parallel
projection using horizontal and vertical collimators between the sample and the detector. 
These projections made from scattered neutrons can be processed into
radiographs and tomographs in the same way as for ordinary neutron or x-ray imaging.
However, these pictures only rely on the scattering cross section. Hence, they provide
unique and complementary information about the sample.
\end{minipage}
~~~~~~~~~~
\begin{minipage}[h]{10cm}
\begin{center}
\epsfig{file=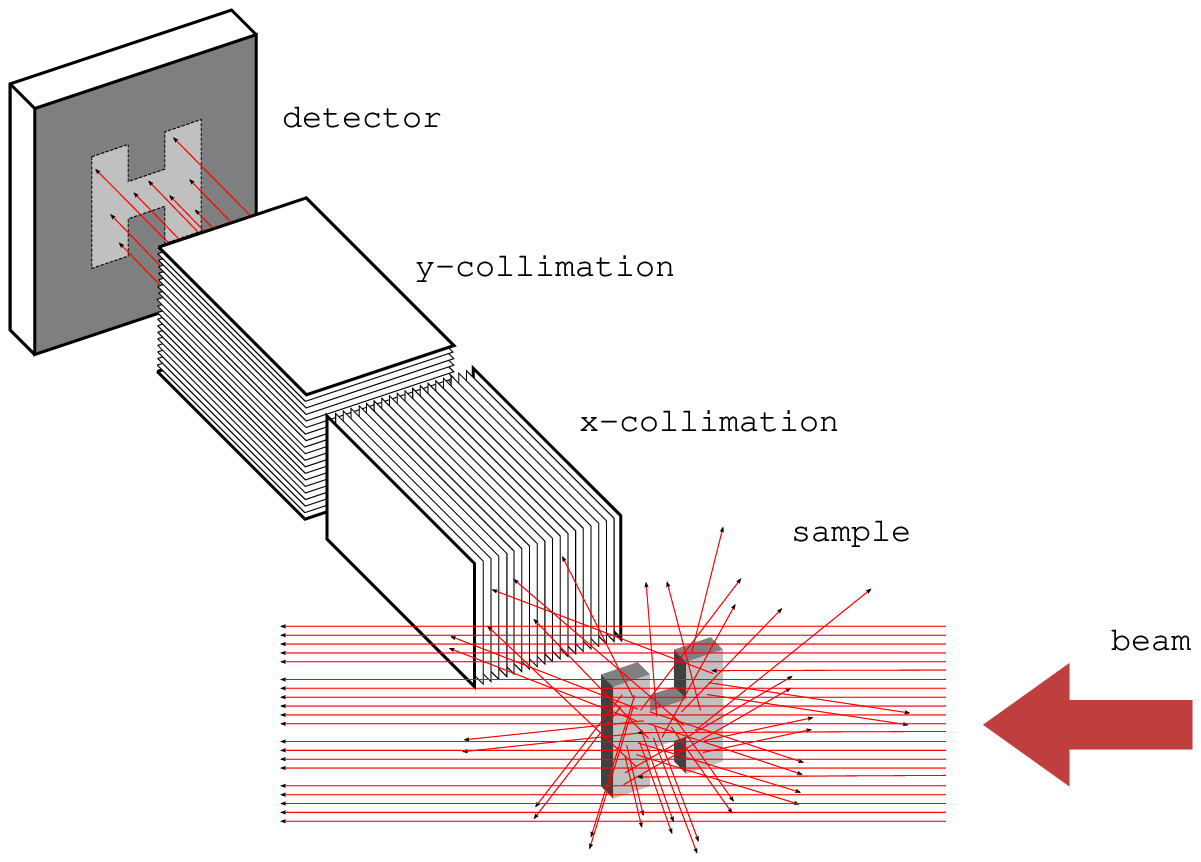,height=5cm}\\
Fig. 2: Scattered neutron imaging: the detector, set far\\
off the beam axis, views the scattered neutrons from the\\
sample via two crossed collimators for position sensitivity.
\end{center}
\end{minipage}

~

\bigskip
In this letter we describe the experimental setup, present first radiographs
and a tomography. We also discuss possible improvements of the setup and 
promising perspectives for the technique of scattered neutron imaging.

\section{Experimental Setup}

The experiment is set up at ILL's neutron tomography station 
\footnote{The neutron tomography station at ILL is a joint test experiment run in collaboration between\\ the
Institut Laue-Langevin and the University of Heidelberg. It has been funded by the German\\ Federal
Ministry for Research and Education under contract number 06HD153I.}. 
Its beamline H9 delivers a thermal flux of
$3\cdot 10^9 \,\textrm{n} \,\textrm{cm}^{-2} \,\textrm{s}^{-1}$
at the sample position.
Three samples of size $3\,\textrm{cm} \times 3 \,\textrm{cm} \times 5 \,\textrm{cm}$ are mounted on a motorized
slide. The are in the form of the letters C, H and L which stand for carbon (graphite), hydrogren (polyethylene) and
lead. The materials were selected because they efficiently scatter neutrons with little absorption:

~

\begin{centering}

	\bigskip
	\begin{tabular}{lccccc}
	\hline\noalign{\smallskip}
	~~~~~~~~~~~~~~~~~~~ & $\sigma_\textrm{abs.} ~ [\textrm{barn}]$ ~~~ & $\sigma_\textrm{coh.} ~ [\textrm{barn}]$ ~~~ & $\sigma_\textrm{inc.} ~ [\textrm{barn}]$ ~~~ & $\sigma_\textrm{tot.} ~ [\textrm{barn}]$ ~~~ & $\mu_\textrm{tot.} ~ [\textrm{cm}^{-1}]$ \\
	\noalign{\smallskip}\hline\noalign{\smallskip}
	hydrogen (PE) & $0.333$ & $1.757$ & $80.26$ & $82.35$ & $\approx 7$ \\
	carbon (graphite) ~~~ & $0.004$ & $5.551$ & $0.001$ & $5.555$ & $0.501$ \\
	lead & $0.171$ & $11.12$ & $0.003$ & $11.29$ & $0.372$ \\
	\noalign{\smallskip}\hline
	\end{tabular}

	\bigskip
	Tab. 1: cross sections and attenuation coefficients
	
\end{centering}

~

\bigskip
The samples are viewed one at a time by two crossed $20^\prime$ Soller collimators.
There is an angle of about $45^\circ$ between the beam axis and the collimator axis.
The collimators are each $40\,\textrm{cm}$ in length, their beam windows are $4\,\textrm{cm}$ wide and divided
into $19$ slits of each about $2 \,\textrm{mm}$. 
The collimated neutrons are detected by a LiF-ZnS-scintillator inside an optics housing. The image is projected
across two Si-Al-surface mirrors onto a $50\,\textrm{mm}$ $f\!=\!1/1.2$ lens and is recorded by
an Andor iXon CCD-camera. 

\bigskip
The detection system is optimized for high speed neutron radiography in the millisecond range
with good statistics. For scattered neutron radiography, the complete setup is carefully
shielded with B$_4$C rubber mats, the more vulnerable parts with
plates of sintered pure $^{10}$B$_4$C. Gamma background is suppressed by
clipping greyvalues above a given threshold. Per projection there are $5$ to $7.5 \cdot 10^{4}$
exposures of each $500\,\textrm{ms}$ to aquire sufficient statistics. A reference image
acquired without any sample in front of the collimators is subtracted from the data images.
No additional digital image filtering is necessary.

~

\bigskip
\section{Experimental Results}

Pyrolithic graphite is well visible in the radiograph (see Fig. 4, left).
Its absorbtion is negligible and it scatters neutrons fairly isotropically
due to its polycristalline structure. The obvious flux reduction on the right
side is due to the attenuation of the beam striking the sample from the left side.

\bigskip
This effect is even more pronounced for hydrogen (see Fig. 4, right),
because its total attenuation coefficient $\mu \approx 7 \, \textrm{cm}^{-1}$
is about fourteen times higher than for graphite.
Consequently, the signal decreases already in the left bar of the letter 'H', which is
one centimeter thick, from left to right; the signal increases in the middle of this bar due to multiple scattering
from below and above; and there is a faint signal on the left side of the right bar, because the front part of this
bar is partly exposed to the unattenuated beam due to the $45^\circ$ orientation of the sample.

\bigskip
In both images the pixelation due to the collimators is visible. It is more pronounced horizontally
than vertically as the $y$-collimator is directly in front of the scintillator. The vertical boundaries
are blurred due to the distance to the $x$-collimator by about the size of a pixel, $2\,\textrm{mm}$,
defining the limit of spatial resolution. The faint bar beneath all letters is that part of the
aluminium sample holder which is within the $4\, \textrm{cm} \times 4 \, \textrm{cm}$ field of view.

\bigskip
Lead is a purely coherent scatterer.
The fact that the letter 'L' is still visible in a scattered neutron radiography (see Fig. 5, left)
is proof of the fact that standard lead is a polycristall. 
However, there are several especially dark spots in the image. At first sight, they could be small yet
strong scatterers embodied in the bulk material.

\newpage
~

\bigskip
\begin{center}

\epsfig{file=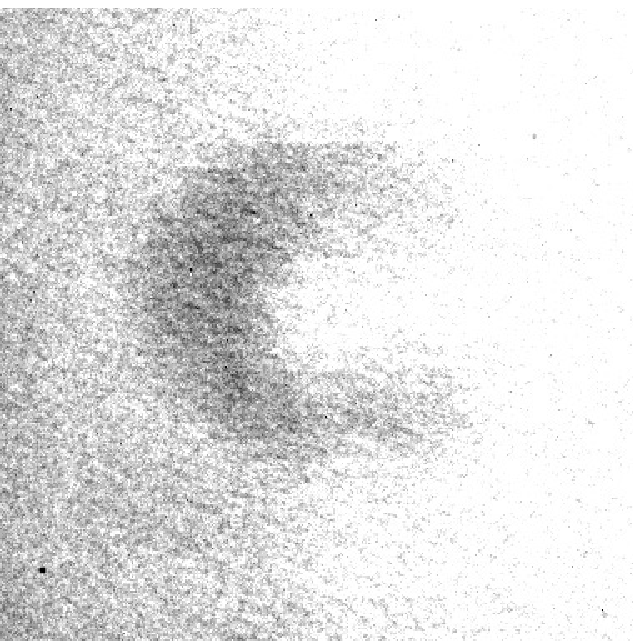,height=6cm}
~~~~~~~~~~
\epsfig{file=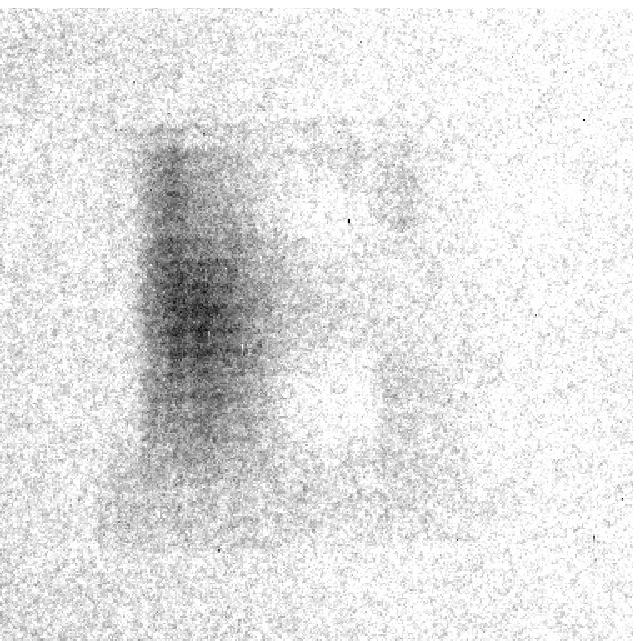,height=6cm}

\bigskip
Fig. 4: scattered neutron radiographies of graphite and polyethylene

~

\bigskip
\epsfig{file=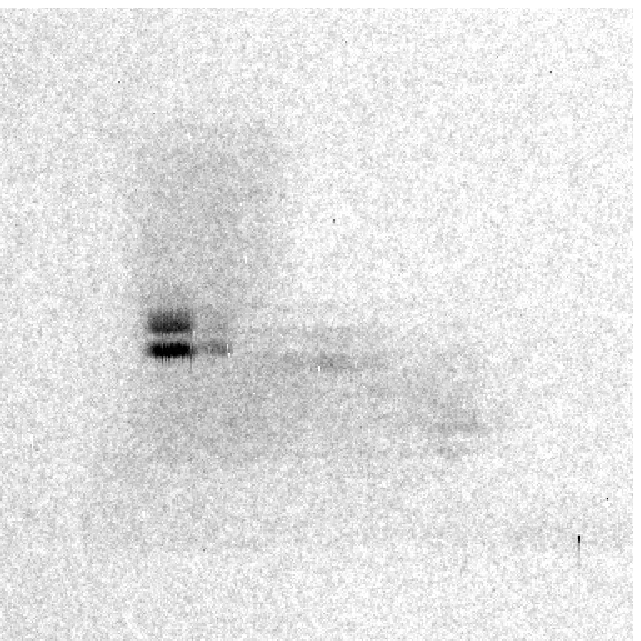,height=6cm}
~~~~~~~~~~
\epsfig{file=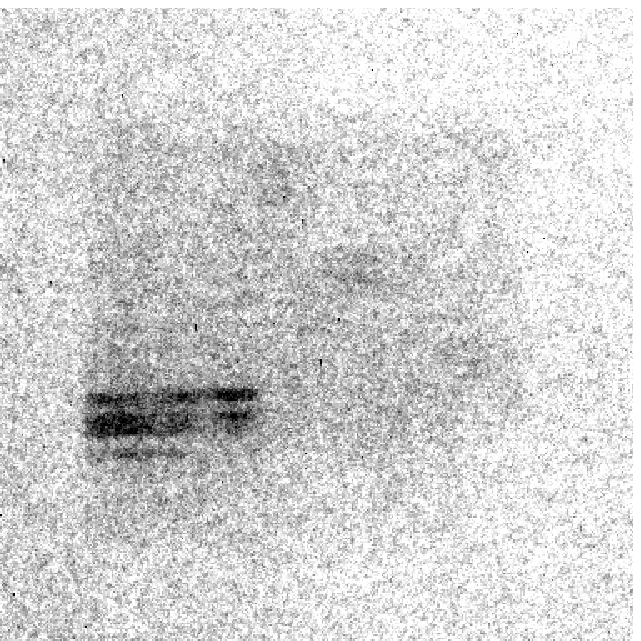,height=6cm}

\bigskip
Fig. 5: (partly coherently) scattered neutron radiographies of lead

~

\bigskip
\begin{minipage}[h]{6.5cm}
\begin{center}

\epsfig{file=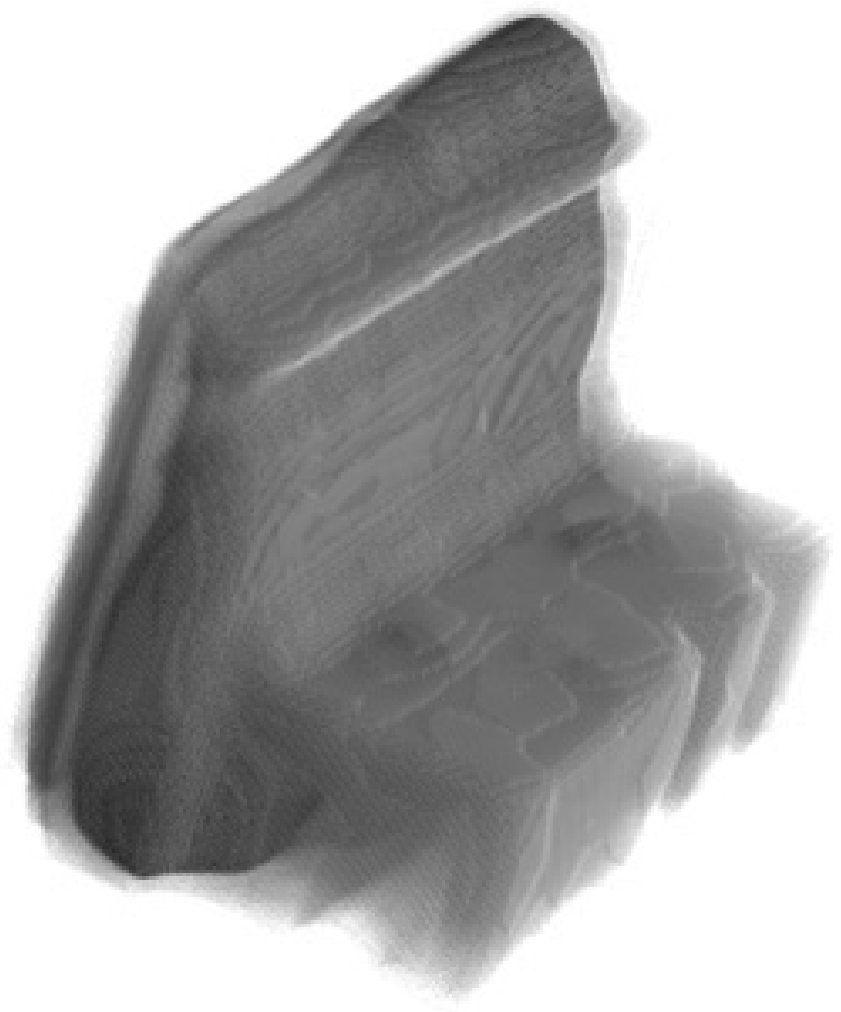,height=6cm}

\bigskip
Fig. 6: scattered neutron tomo-\\graphy of the lead sample

\end{center}
\end{minipage}
~~~~~~~
\begin{minipage}[h]{6.5cm}
\begin{center}

\epsfig{file=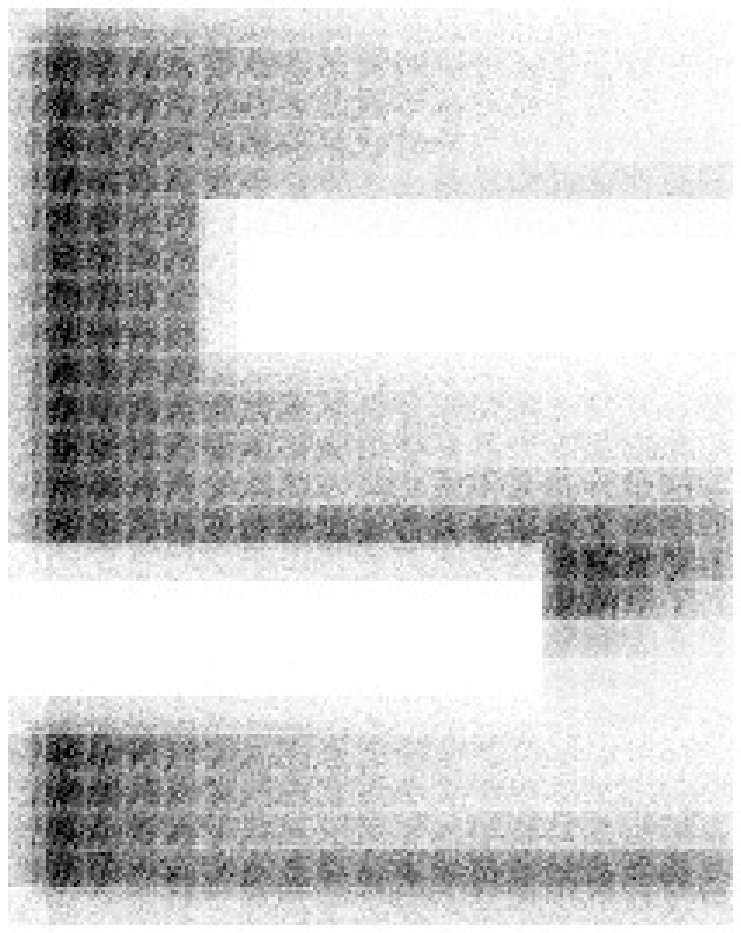,height=6cm}

\bigskip
Fig. 7: Monte Carlo simulation\\displaying effects of self-attenuation

\end{center}
\end{minipage}

\end{center}

\newpage

\bigskip
To check this hypothesis the sample was turned by $90^\circ$ relative to the detector.
The image, now showing the back of the letter 'L' (see Fig. 5, right), was combined from two exposures
for a larger field of view. Again, there appear peaks in the signal. This time, however, they show up
at a different height in the sample and can thus not originate from the same region in the sample.
The conclusion is that the peaks in the signal are produced by small coherently scattering crystals
orientated by chance to fulfill the Bragg condition in direction of the detector. When the sample
is rotated, different scattering centers flash up at different orientations.

\bigskip
Several exposures of the same image, for example of the letter 'L', can be processed into a scattered
neutron tomography of the sample (see Fig. 6). Especially for this purpose, the effects of coherent
scattering and self-attenuation within the sample, need to be taken into account in the future.
To study these effects systematically, Monte Carlo simulations are being developed (see Fig. 7).

\bigskip
\section{Conclusion and Outlook}

Images by scattered neutron radiography and tomography are presented.
Both incoherent (H) and coherent scatterers (C, Pb), and even weak
scatterers (Al) are visualized with millimeter resolution.
It is worthwhile to consider possible improvements of the method:

\bigskip
Using crossed collimators, the resolution is limited by the width of the slits.
Here flux can be gained by making the collimators just shorter, without any
loss of position sensitivity. With collimators of $4\, \textrm{cm}$ in
length instead of $40\, \textrm{cm}$, a factor of $100$ in flux can be gained.
The limit is the size of the sample, which should be smaller than the
collimators because of the blurring of the projection. A real improvement
is a honeycomb collimator which combines $x$- and $y$-collimation on the
same length as a Soller collimator, doubling resolution and quadrupling
flux.


\bigskip
In contrast to ordinary neutron imaging, image quality and resolution are not
dependent on the collimation of the incoming beam. This is a strong point in
favour of high-flux medium-collimation installations like at ILL. It would
be possible to set up an instrument dedicated to scattered neutron imaging very close to a reactor
core or a spallation target, for example at an existing irradiation facility.
Also, several detector patches can be positioned around the sample forming an
effective $4\pi$-detector.

\bigskip
Scattered neutron imaging adds another geometrical degree of freedom to
neutron imaging. Coherently scattering domains will lighten up under Bragg
angles, so their three dimensional distribution and orientation in the
sample can be determined for 3D-crystallography. On the other hand, and more fundamentally,
scattered neutron imaging adds two spatial degrees of freedom to general
neutron spectrometry, diffractometry, etc. In principle, regardless of flux
and resolution, the hallmark experiments of neutron science could broaden
their horizon from small homogenous samples to structured heterogenous objects.

\end{document}